\documentclass[epjST]{svjour}
\usepackage{graphics}
\usepackage{color}
\usepackage{amsmath,amssymb}
\newcommand{\nmax}{n_{\mathrm{max}}} 
\newcommand{\peq}{p^{\mathrm{eq}}_n(\mu)}
\newcommand{\nav}{\langle n \rangle}
\newcommand{\nvar}{\langle n^2 \rangle - \langle n \rangle^2}
\begin{document}
\title{Adsorption and desorption in confined geometries: a discrete hopping model}
%\subtitle{}
\author{T. Becker\inst{1}\fnmsep\thanks{\email{thijsbecker@gmail.com}} \and K. Nelissen\inst{2,1} \and B. Cleuren\inst{1} \and B. Partoens\inst{2} \and C. Van den Broeck\inst{1}}
\institute{Hasselt University, B-3590 Diepenbeek, Belgium \and Departement Fysica, Universiteit Antwerpen, Groenenborgerlaan 171, B-2020 Antwerpen, Belgium}
\abstract{
We study the adsorption and desorption kinetics of interacting particles moving on a one-dimensional lattice. Confinement is introduced by limiting the number of particles on a lattice site. Adsorption and desorption are found to proceed at different rates, and are strongly influenced by the concentration-dependent transport diffusion. Analytical solutions for the transport and self-diffusion are given for systems of length 1 and 2 and for a zero-range process. In the last situation the self- and transport diffusion can be calculated analytically for any length.  
} 
\maketitle
\section{\label{intro}Introduction}

Diffusion in confined geometries is ubiquitous in nature, for example in biological cells \cite{brangwynne2009intracellular,MuthukumarPNAS}. An important case is diffusion in microporous materials, such as zeolites and metal-organic frameworks (MOFs) \cite{book_diffnano}. Because of their structure on the molecular scale and large surface area, they are ideally suited for e.g.~catalysis and particle separation.
Microporous materials occur naturally, but can also be made in the laboratory. New fabrication techniques have led to a large increase in available materials, with a great diversity in possible structures \cite{yaghi2003reticular,porousapp,ACSnano}. 
Many applications require a proper understanding of how particles diffuse in these materials. Thanks to recent advances a detailed experimental view of diffusion in microporous materials is now available \cite{Naturekarger2014}. A theoretical analysis is notoriously difficult due to the complex interactions involved. Progress has been made by advanced molecular dynamics simulations \cite{PRLbeerdsen2006,smit2008molecular,CSRkrishna2012}, or via coarse-grained stochastic models in continuous space \cite{zwanzig1992diffusion,PRErubi2001,Burada2009,PREkalinay,EPJSTholek,PREcarvalho} or on a lattice \cite{SurfScireed1981,IRPCauerbach}. In a recent paper \cite{PRLbecker} we introduced a lattice model that provides an intuitive interpretation of the role of interactions upon the transport and self-diffusion.
By fitting only equilibrium properties, good agreement was found with experimental results of methanol diffusion in MOF ZIF-8 \cite{PRLchmelik2010}.

In this work, we discuss the adsorption and desorption kinetics of this model. The rates at which particles are absorbed/desorbed from the material are of crucial importance for many applications \cite{BOOKruthven}. Particle interactions have a large influence on the adsorption and desorption behavior \cite{JACStsotsalas2013}. The influence of interactions is therefore studied in detail.

The paper is organized as follows. We introduce the model in Section \ref{sec:2}. In Section \ref{sec:3} we present analytical expressions of the transport and self-diffusion coefficients. Transport diffusion plays a crucial role here, since adsorption/desorption is the result of mass transfer in response to a concentration gradient. For certain parameters the model reduces to a zero-range process. In this case it is always analytically solvable. In Section \ref{sec:4} the adsorption and desorption kinetics of the model is discussed. Our main conclusions are presented in Section \ref{sec:conclude}.

\section{\label{sec:2}The model}

\begin{figure}
\centering
\resizebox{0.7\columnwidth}{!}{%
\includegraphics{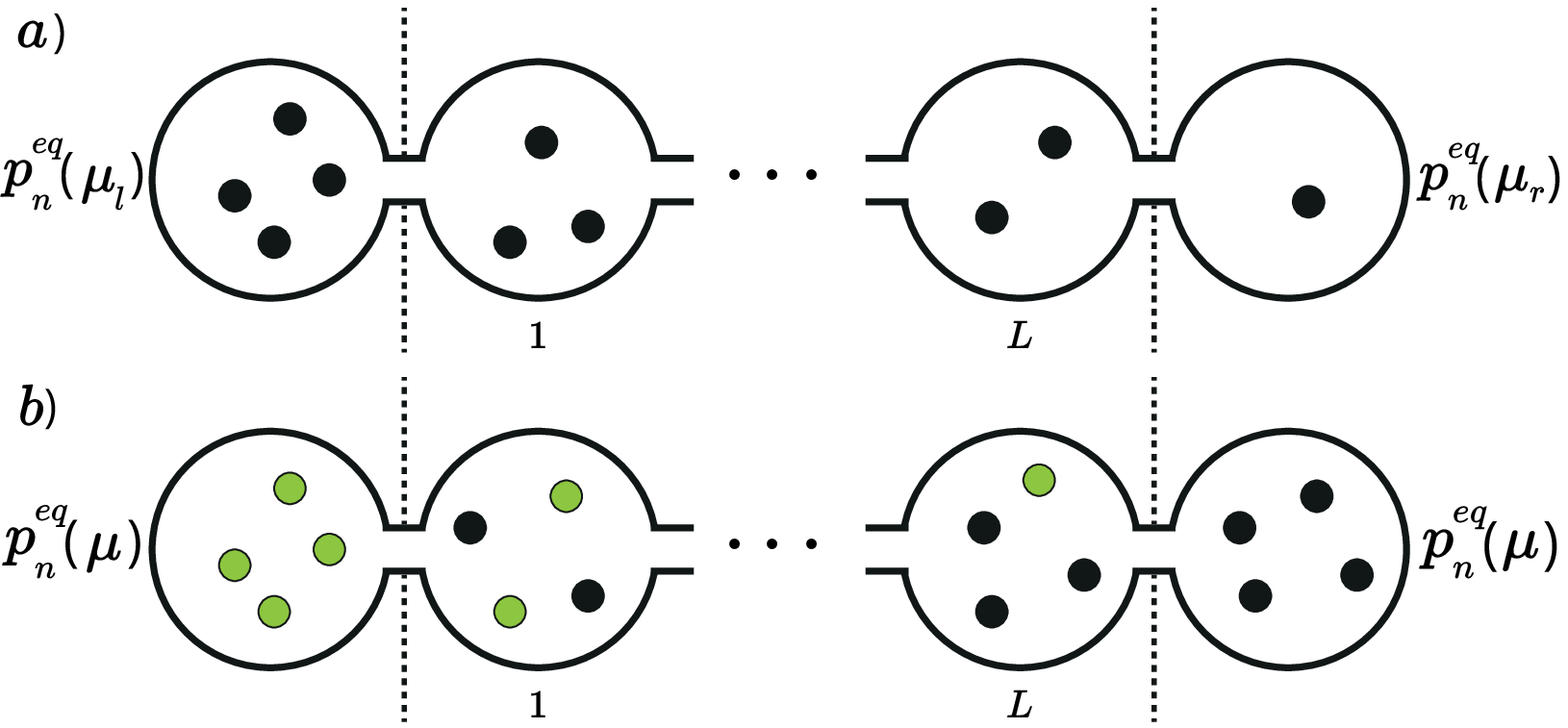} }
\caption{The system, shown between dotted lines, consists of cavities connected by narrow windows. On the boundaries the system is connected to cavities with an uncorrelated equilibrium distribution. $a)$ The transport diffusion is measured under a steady concentration gradient $\mu_l \neq \mu_r$, in first order around equilibrium. $b)$ The self-diffusion is measured in equilibrium ($\mu_l = \mu_r = \mu$), where a concentration gradient of labeled (green) particles is introduced.}
\label{fig:1}
\end{figure}

The system we consider is a one-dimensional array of $L$ cavities, each one connected to its nearest neighbours. The outer left and right cavities are connected to reservoirs, allowing particles to enter/leave the system. A sketch is shown in Fig.~\ref{fig:1}. Particles can move between the cavities by passing through the narrow channel in between. This type of system is a paradigmatic model for the study of diffusion under confinement \cite{PRLreguera2006,PRLghosh,JCPmalgaretti2013,PRLmartens,EPJSTmartens}.
Due to the narrow passages, the time spent by the particles in a cavity before moving to one of its neighbours is considered much larger than the relaxation time inside a cavity. This is a common assumption in the modeling of such structures \cite{PRLbeerdsen2004,CESTunca2003,NPSruthven1971,JCPpazzona20091,JCPpazzona20092}. Consequently, this separation of time scales allows for a coarse graining of the intracavity degrees of freedom \cite{PREesposito2012}, and the state of cavity $i \in \{1,L\}$ is characterized solely by the number of particles $n_i$ it contains. The interactions of $n$ particles in a cavity are described by the equilibrium free energy
\begin{equation}
F(n) = U(n) - T S(n),
\end{equation}
with $U(n)$ and $S(n)$ the average energy and entropy respectively. The system is at constant temperature $T$.
When the system is in equilibrium with a reservoir at chemical potential $\mu$, the probability to find $n$ particles in any cavity is:
\begin{equation}\label{eq:peq}
p^{\mathrm{eq}}_n (\mu) = \mathcal{Z}(\mu)^{-1}e^{- \beta [F(n) - \mu n]},
\end{equation}
with $\mathcal{Z}(\mu)$ the normalization constant.
Contributions to $F(n)$ come from particle-particle and particle-cavity interactions. We assume that there is no interaction between particles residing in different cavities. This is a good quantitative approximation for low and medium particle concentrations \cite{CESTunca2003,JPCCpazzona2012}. For high particle concentrations such an approximation can, in general, only be expected to lead to a qualitative agreement with experimental systems.

Since our objective is to study the influence of the confinement and interaction upon the adsorption/desorption behaviour, we decompose $F(n)$ as follows:
\begin{equation}
F(n) = F^{\mathrm{id}}(n) + f(n),
\end{equation}
where $F^{\mathrm{id}}(n)$ is the free energy of an ideal gas of $n$ indistinguishable particles in a cavity of volume $V$:
\begin{equation}\label{eq::Fidealgas}
F^{\mathrm{id}}(n) \equiv k_B T \left[ \ln (n!) - n \lnÊ\left( V / \Lambda^3  \right) \right],
\end{equation}
with $\Lambda = h /  \sqrt{2 \pi m k_B T}$ the thermal de Broglie wavelength, $m$ the mass of one particle, $h$ the Planck constant, and $k_B$ the Boltzmann constant. Note that a linear term in $F(n)$ simply rescales the chemical potential of the system Eq.~\eqref{eq:peq}. Such a linear term does therefore not influence the equilibrium statistics at a given particle concentration.
The various interactions are included in $f(n)$, called the \textit{interaction free energy}. An important contribution comes from the finite volume of the cavities, which limits the amount of particles inside a single cavity. The maximum number of particles allowed simultaneously in a cavity is denoted by $\nmax$, which implies that $f(\nmax+1)$ diverges. Confinement is also present in $f(n)$, which is influenced by excluded volume interactions between the particles and the interaction with the cavity wall.

Due to the coarse graining, the complete state of the system is specified by the number of particles in each cavity $(n_1,n_2,\ldots,n_L)$. This state changes as particles move to neighboring cavities or when particles enter/leave the system via the boundaries. The dynamics is Markovian. The rate for a particle to jump from a cavity containing $n$ particles to a cavity containing $m$ particles is denoted by $k_{nm}$ and is required to satisfy the detailed balance condition:
\begin{equation}
k_{n m} p^{\mathrm{eq}}_n(\mu) p^{\mathrm{eq}}_m(\mu) = k_{m+1,n-1} p^{\mathrm{eq}}_{n-1}(\mu) p^{\mathrm{eq}}_{m+1}(\mu).
\end{equation}
A particular choice of rates which satisfy this condition is
\begin{equation}\label{eq:ratesletter}
k_{nm} = n e^{-(\beta / 2) \left[ f(n-1) + f(m+1) - f(n) - f(m) \right]}.
\end{equation}
Section \ref{sec:ZRP} considers an alternative choice. Particles enter or leave the system via the left and right reservoirs at chemical potential $\mu_l$ and $\mu_r$ respectively. These reservoirs are modeled as cavities whose state is uncorrelated from the system and is given by the equilibrium probability distributions $p^{\mathrm{eq}}_n(\mu_l)$ and $p^{\mathrm{eq}}_n(\mu_r)$. The rates at which a reservoir cavity at chemical potential $\mu$ adds (index $+$) or removes (index $-$) a particle from a cavity containing $n$ particles are given by:
\begin{align}\label{eq:trb}
k^+_n(\mu)  = \sum_m k_{mn} p^{\mathrm{eq}}_m(\mu)\;\;\; ; \;\;\; k^-_n(\mu) = \sum_m k_{nm} p^{\mathrm{eq}}_m(\mu).
\end{align}
With the transition rates determined, one can setup the master equation describing the time evolution of the probability $p_{n_1,n_2,\ldots,n_L}(t)$ for the system to be in state $(n_1,n_2,\ldots,n_L)$ at time $t$.
%%%%%%%%%%%%%%%%%%%%%%%%%%%%%%%%%%%%%%%%%%%%%%%%%%%%%%%%%%%%%%%%%%
\section{\label{sec:3}Analytical solutions for the diffusion}
%%%%%%%%%%%%%%%%%%%%%%%%%%%%%%%%%%%%%%%%%%%%%%%%%%%%%%%%%%%%%%%%%%
Diffusion of particles in these systems can be described by different diffusion coefficients. For adsorption and desorption processes, which are the result of a concentration gradient, the coefficient of interest is the transport diffusion $D_t$ \cite{book_diffnano}. This coefficient characterizes the linear response of the system via Fick's first law:
\begin{equation}
j = - D_t \frac{\partial c}{\partial x},
\end{equation}
where $j$ is the particle flux and $c$ the concentration of particles. For interacting particles, the transport diffusion is in general different from the self-diffusion coefficient $D_s$ \cite{PRLbecker,PRLchmelik2010}, which measures the average mean square displacement (MSD) of a single particle in equilibrium. In one dimension it is defined as:
\begin{equation}\label{eq:selfdiff1}
D_s = \lim_{t \uparrow \infty} \frac{1}{2 t} \overline{ \left[ x(t) - x(0) \right]^2 },
\end{equation}
where $x(t)$ is the position of the particle at time $t$ and the overline denotes the average over all trajectories. An alternative method for calculating $D_s$ is shown in Fig.~\ref{fig:1}$(b)$. Particles in the left reservoir are labeled, resulting in a concentration gradient of labeled particles throughout the system. The label is introduced for monitoring purposes only, and does not change the physical properties of the particles. The self-diffusion as defined in Eq.~(\ref{eq:selfdiff1}) is equivalent to \cite{BOOKkarger}:
\begin{equation}\label{eq:selfdiff2}
j^* = - D_s \frac{\partial c^*}{\partial x},
\end{equation}
with $j^*$ and $c^*$ respectively the flux and the concentration of labeled particles, under overall equilibrium conditions ($\mu_l = \mu_r = \mu$).
%%%%%%%%%%%
\subsection{Systems of length $L = 1$}\label{sec:nocorr} 
For a system of length $L=1$ the calculation of both transport and self-diffusion can be done analytically. The probability $p_n(t)$ to find $n$ particles inside the cavity satisfies the master equation
\begin{equation}
\dot{p}_n(t)=-(k_n^++k_n^-)p_n(t)+k_{n-1}^+p_{n-1}(t)+k_{n+1}^-p_{n+1}(t),
\end{equation}
with $k_n^{\pm}=k_n^{\pm}(\mu_l)+k_n^{\pm}(\mu_r)$. The stationary solution is easily obtained and reads:
\begin{equation}
p_n=\frac{k_{n-1}^+\ldots k_1^+k_0^+}{k_n^-\ldots k_2^-k_1^-}p_0,
\end{equation}
with $p_0$ determined through normalization. The flux of particles is measured between the left reservoir and the cavity:
\begin{equation}
j=\sum_n\left[k_n^+(\mu_l)-k_n^-(\mu_l)\right]p_n.
\end{equation}
The concentration gradient is related to the difference in particle numbers and reads:
\begin{equation}
\frac{\partial c}{\partial x} = \frac{1}{\lambda^2}\sum_{n, n_l} \left( n - n_l \right) p^{\mathrm{eq}}_{n_l}(\mu_l) p_{n},
\end{equation}
where $\lambda$ is the center-to-center distance between two neighboring cavities. By definition, the transport diffusion quantifies the linear response of the flux with respect to the concentration gradient. Hence we make an expansion to first order in $\delta = (\mu_l - \mu_r)/2$. The stationary distribution becomes $p_n = p^{\mathrm{eq}}_{n}(\mu) + O(\delta^2)$ with $\mu = (\mu_l + \mu_r)/2$. The resulting transport diffusion reads \cite{PRLbecker}:
\begin{equation}
D_t=\frac{\lambda^2\sum_{n,m}k_{nm}p^{\mathrm{eq}}_{n}(\mu)p^{\mathrm{eq}}_{m}(\mu)}{\langle n^2 \rangle - \langle n \rangle^2}= \frac{\lambda^2\langle k \rangle}{\nvar}, \label{eq:dtnocorr}
\end{equation}
where $\langle \cdot \rangle$ refers to the average over the equilibrium distribution Eq.~\eqref{eq:peq}.

The self-diffusion coefficient $D_s$ is calculated via the alternative method. Knowledge of the total number of particles inside the cavity $n$ is not sufficient to completely specify the state of the cavity, but has to be supplemented by the number $n^*$ of labeled particles. The probability for state $(n,n^*)$ is denoted by $p_{n\vert n^*}$. All particles coming from the left reservoir are labeled, those from the right reservoir are not labeled. For this situation the stationary distribution reads \cite{PRLbecker}:
\begin{equation}
p_{n\vert n^*}=p^{\mathrm{eq}}_{n}(\mu)\frac{1}{2^n}\frac{n!}{n^*!(n-n^*)!}.
\end{equation}
This result is exact and does not require any expansion since the chemical potential in the left and right reservoirs are equal and set to $\mu$. The flux and concentration gradient of labeled particles between the left reservoir and cavity are:
\begin{eqnarray}
j^*&=&\sum_{n_l,n,n^*}\left(k_{n_ln}-k_{Ðnn_l}\frac{n^*}{n} \right)p_{n\vert n^*}p^{\mathrm{eq}}_{n_l}(\mu)=\frac{\langle k \rangle}{2}, \\
\frac{\partial c^*}{\partial x}&=&\frac{1}{\lambda^2}\sum_{n_l,n,n^*}(n^*-n_l)p_{n \vert n^*}p^{\mathrm{eq}}_{n_l}(\mu)=-\frac{\langle n \rangle}{2\lambda^2}.
\end{eqnarray}
And hence
\begin{equation}
D_s(\mu) = \lambda^2 \frac{\langle k \rangle}{\langle n \rangle}. \label{eq:dsnocorr}
\end{equation}
Since the system is uncorrelated with the reservoirs and contains only 1 cavity, the results Eqs.~\eqref{eq:dtnocorr} and \eqref{eq:dsnocorr} do not include any effects of correlations. In fact, these results are valid for any system length, provided correlations are neglected \cite{Arxivbecker}. The explanation goes as follows. Suppose we want to measure the self-diffusion by calculating the average MSD of a single tagged particle. Subsequent jumps of the tagged particle are correlated because of memory effects in the environment. A well known example is the back-correlation mechanism in systems where each lattice site can hold at most one particle. In this case the particle is more likely to jump back, because the site it jumped from is more likely to be empty. If the environment is memoryless the particle always sees neighboring cavities with distribution $\peq$. As a result subsequent particle jumps are uncorrelated. It can be shown that, in this situation, the transport and self-diffusion are given by respectively Eqs.~\eqref{eq:dtnocorr} and \eqref{eq:dsnocorr} \cite{Arxivbecker}. 
One finds for $L=1$:
\begin{equation}\label{eq:thdf}
\frac{D_t(\mu)}{D_s(\mu)} = \frac{\nav}{\nvar} \equiv \Gamma(\mu),
\end{equation}
where $\Gamma(\mu)$ is called the thermodynamic factor. If one neglects all correlations Eq.~(\ref{eq:thdf}) can be derived from a general argument \cite{RPPgomer1990}. It is not valid if correlations are included.

\subsection{\label{sec:L2}Systems of length $L=2$ and $\nmax = 2$}
\begin{figure}
\centering
\resizebox{0.8\columnwidth}{!}{%
\includegraphics{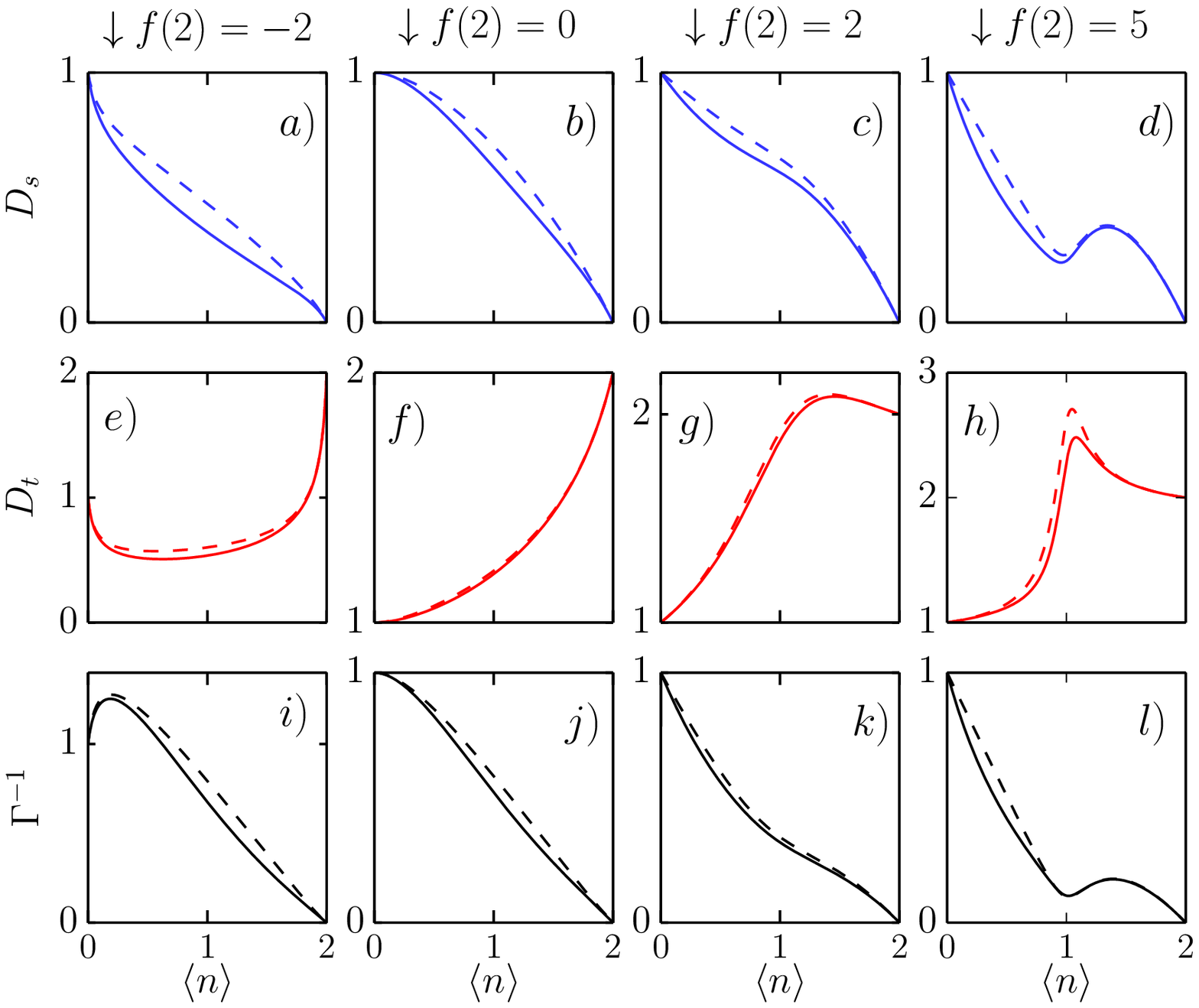} }
\caption{System of length $L = 2$, $\nmax = 2$, and rates Eq.~(\ref{eq:ratesletter}). $a)$, $b)$, $c)$, and $d)$: self-diffusion $D_s$, exact (solid line) and uncorrelated (dashed line); $e)$, $f)$, $g)$, and $h)$: transport diffusion $D_t$, exact (solid line) and uncorrelated (dashed line); $i)$, $j)$, $k)$, and $l)$: $\Gamma^{-1}$ (dashed line) and $D_s/D_t$ (solid line), for respectively $f(2) = -2$, $f(2) = 0$, $f(2) = 2$, and $f(2) = 5$.}
\label{fig:2}
\end{figure}
The influence of correlations appear only for system sizes $L=2$ and larger. However, for increasing system size $L$ and $\nmax$ the resulting calculations quickly become unfeasible, even when making use of a symbolic calculator. We were able to calculate analytically the self- and transport diffusion for systems of length $L=2$ and $\nmax = 2$, for the rates Eq.~(\ref{eq:ratesletter}). As follows from Eq.~(\ref{eq:ratesletter}), adding a linear term to $f(n)$ does not influence the dynamics. As discussed in Section \ref{sec:2}, the equilibrium statistics in function of particle concentration is also not influenced by a linear term in $f(n)$. Without loss of generality, we rescale the interaction free energy by $f(n) \rightarrow f(n) - nÊ[ f(1) - f(0) ] - f(0)$, which makes $f(0) = f(1) = 0$. Both the equilibrium and dynamical quantities then only depend on $f(2)$.

The exact and uncorrelated results for $f(2) = -2, 0, 2,$ and $5$ are plotted in Fig.~\ref{fig:2}. $D_s$ has a minimum at $\nav = 1$ for $f(2) = 5$ because the state $(1,1)$ is very stable, and particles will not diffuse easily. The transport diffusion shows a maximum in this situation, because a particle that enters the system when it is in state $(1,1)$ is ``pushed out'' again rapidly. The transport diffusion has a minimum for low and medium concentrations for $f(2) = -2$ because particles are attracted against the concentration gradient.

Correlations always lower the self-diffusion compared to the uncorrelated result. This was checked analytically for all interactions. Correlations lower the transport diffusion almost always compared to the uncorrelated result, except for $\ln (2) < f(2) < \ln (9/4)$. For these interactions the transport diffusion is slightly higher than the uncorrelated result. The particle concentrations at which this occurs depends on the interaction. Because the effect is very small we do not plot this situation. It is an open question whether positive correlations for the transport diffusion also exist for systems with $L \uparrow \infty$, when the influence of the reservoir cavities is negligible.

%%%%%%%%%%%%%%%%%%%%%%%%%%%%%%%%%
\subsection{Zero-range processes}\label{sec:ZRP}
Exact expressions for transport and self-diffusion are possible when the model satisfies the criteria of a zero-range process (ZRP) \cite{ZRPevansreview}. In a ZRP the transition rates only depend on the number of particles in the departing cavity, i.e. the rates must be of the form $k_{nm} = k_n$. Hence, the particle is not aware of the state of the cavity it moves to. Clearly this can only be true if the number of particles at each site is unlimited, i.e.~$\nmax = \infty$. Local detailed balance then gives the following condition:
\begin{equation}
k_n \frac{\peq}{p^{\mathrm{eq}}_{n-1}(\mu)} = k_{m+1} \frac{p^{\mathrm{eq}}_{m+1}(\mu)}{p^{\mathrm{eq}}_{m}(\mu)}.
\end{equation}
Since this condition must hold for all values of $m$ and $n$, both sides of the equation must be equal to a function $g(\mu)$. Using Eq.~(\ref{eq:peq}) one finds:
\begin{equation}
k_n = g(\mu) e^{-Ê\beta \mu} n e^{\beta [ f(n) - f(n-1) ]}.
\end{equation}
Since $k_n$ only depends on particle interactions in the cavity it must be independent of the chemical potential of the reservoir. As a result $g(\mu) = \nu e^{\beta \mu}$ with $\nu$ a positive constant. The model is therefore a ZRP for the rates:
\begin{equation}\label{eq:trzrp}
k_{n} = \nu n e^{\beta [ f(n) - f(n-1)Ê]}.
\end{equation}
As before, a linear term in the free energy is not relevant, and can be absorbed in $\nu$. The transition rates describing the exchange of particles with the reservoirs follow directly by substituting Eq.~\eqref{eq:trzrp} in Eq.~\eqref{eq:trb}:
\begin{align}
k^+_n(\mu)  = \nu e^{\beta \mu}\;\;\; ; \;\;\; k^-_n(\mu) = k_n.
\end{align}
For a ZRP $p_{n_1,n_2,\ldots,n_L}(\mu_l,\mu_r)$ can be calculated exactly:
\begin{equation}\label{eq::ZRPtranssol}
p_{n_1,n_2,\ldots,n_L}(\mu_l,\mu_r) = \prod_{i=1}^L p^{\mathrm{eq}}_{n_i}(\mu_i),
\end{equation}
with
\begin{equation}
\mu_i = \beta^{-1} \ln \left[ e^{\beta \mu_l} - \frac{i}{L+1} \left( e^{\beta \mu_l} - e^{\beta \mu_r} \right)  \right].
\end{equation}
It is important to note here that it is not always possible for the system to reach a stationary state. For example attractive particles can condensate in the system, which continues to absorb particles from the reservoirs since $\nmax = \infty$. We refer to \cite{ZRPlevinesolution} for a derivation of Eq.~(\ref{eq::ZRPtranssol}) and a discussion on its range of validity. This type of interactions are excluded in the following discussion. The solution is a product measure: particle numbers in different cavities are uncorrelated, for all possible interactions. Performing a first order expansion in $\delta$ around $\delta = 0$, as explained in Section \ref{sec:2}, one finds that $\mu_i$ decreases linearly between the cavities:
\begin{equation}
\mu_i = \mu + \delta \left( 1 - \frac{2 i}{L+1}   \right).
\end{equation}
From this result one can derive that $D_t$ is equal to the uncorrelated result Eq.~\eqref{eq:dtnocorr}. 

The self-diffusion coefficient of a ZRP can be calculated directly from the definition Eq.~\eqref{eq:selfdiff1}, see for example \cite{JSPasselah1997}. We succeeded to calculate $D_s$ via the alternative method by introducing labeled particles in the system. The stationary solution is again a product measure:
\begin{equation}\label{eq::sollabelZRP}
p_{n_1, n^*_1, \ldots, n_L, n^*_L}(\mu,\alpha_{l}, \alpha_{r}) = \prod_{i=1}^L p^{\mathrm{eq}}_{n_i}(\mu) \binom{n_i}{n_i^*} \alpha_i^{n_i^*} (1 - \alpha_i)^{n_i-n_i^*},
\end{equation} 
with $\alpha_{l}$ and $\alpha_{r}$ the fraction of labeled particles in the left and right reservoir respectively and
\begin{align}\label{eq::alphanc} 
\alpha_i = \alpha_{l} - (\alpha_{l} - \alpha_{r}) \frac{i}{L+1}.
\end{align}
The fraction of labeled particles decreases linearly between the cavities. From Eq.~(\ref{eq::sollabelZRP}) one can derive that $D_s$ is equal to the uncorrelated result Eq.~(\ref{eq:dsnocorr}). 
%%%%%%%%%%%%%%%%%%%%%%%%%%%%%%%%%%%%%%%%%%%%%%%%%%%%%%%%%%%%%%%%%%
\section{Adsorption and desorption kinetics}\label{sec:4}
%%%%%%%%%%%%%%%%%%%%%%%%%%%%%%%%%%%%%%%%%%%%%%%%%%%%%%%%%%%%%%%%%%
In this Section we investigate the adsorption and desorption kinetics of the model. 
A one-dimensional system of length 100 is considered. We assume that there is no extra resistance at the boundaries, i.e.~there are no surface barriers, which can be the case in experiments \cite{CITchmelik2010}. The process is assumed to be isothermal, which is a good approximation for materials of small size.
The dynamics is simulated using kinetic Monte Carlo (kMC), see the supplementary material of \cite{PRLbecker}. Adsorption and desorption runs are performed between $5.10^3$ and $3.10^4$ times each (depending on the interaction) to achieve good statistics. We set $\lambda = \nu = 1$ in the simulations.
%%%%%%%%%%%
\subsection{Non-interacting particles}

\begin{figure}
\centering
\resizebox{1.0\columnwidth}{!}{%
\includegraphics{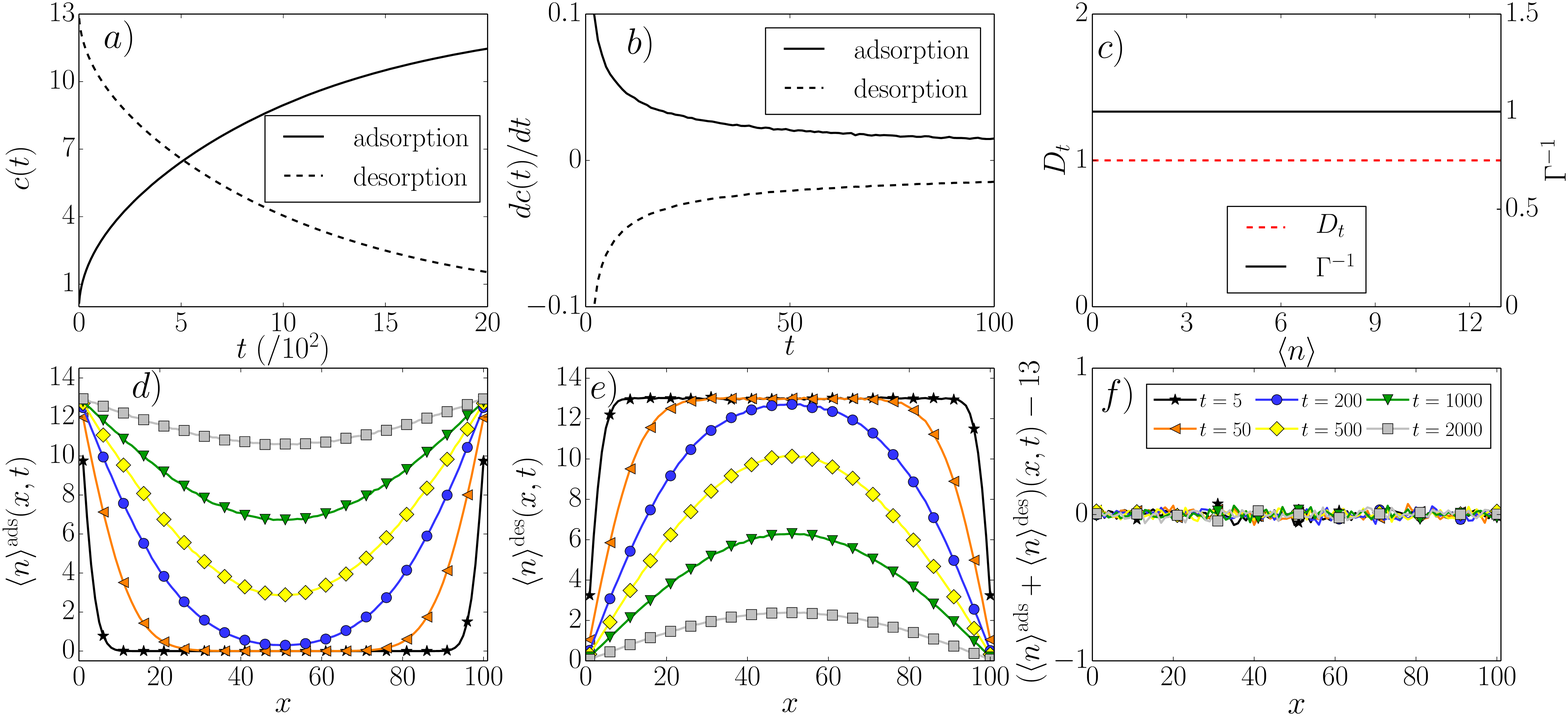} }
\caption{Adsorption/desorption between $\nav = 13$ and $\nav = 0$, for $f(n) =0$, no $\nmax$, and rates $k_n = n$. $a)$ Average concentration $c(t)$. $b)$ Rate of adsorption and desorption. $c)$ Transport diffusion (analytical solution) and $\Gamma^{-1}$. $d), e)$ Average number of particles $\nav(x,t)$ in each cavity at different times $t$, during respectively adsorption $\nav^{\mathrm{ads}}(x,t)$ and desorption $\nav^{\mathrm{des}}(x,t)$. For visual clarity markers are shown each 5 positions. The lines are a guide to the eye. $f)$ $\nav^{\mathrm{ads}}(x,t) +\nav^{\mathrm{des}}(x,t) - 13$. Markers are shown each 10 points. The lines are a guide to the eye.}
\label{fig:adsnoint}
\end{figure}

In Fig.~\ref{fig:adsnoint} we plot the adsorption and desorption kinetics for non-interacting particles $f(n) = 0$ and no $\nmax$. Both rates Eqs.~(\ref{eq:ratesletter}) and (\ref{eq:trzrp}) are equal to $k_{nm} = \nu n$, and $\peq$ is the Poisson distribution with average $\nav$, for which $\langle n^2 \rangle - \nav^2 = \nav$. Since all particle jumps are uncorrelated $D_t$ is given by Eq.~(\ref{eq:dtnocorr}), which reduces to $\nu \lambda^2$, cf.~Fig.~\ref{fig:adsnoint}(c). 

In the desorption process the system is equilibrated according to $\peq$ with the chemical potential corresponding to $\nav = 13$. Starting at time $t=0$ the reservoir cavities are put at chemical potential $\mu \rightarrow - \infty$ for all times ($p^{\mathrm{eq}}_n(-\infty) = \delta_{n0}$). The adsorption proceeds oppositely: the system starts in a completely empty state, and the reservoir cavities are put at the chemical potential corresponding to $\nav = 13$ at time $t = 0$.
The average number of particles in cavity $x$ at time $t$ is denoted by $\nav (x,t)$. The average particle concentration in the system, $c(t) = \sum_{x=1}^{L} \nav(x,t) / L$, during adsorption and desorption is shown in Fig.~\ref{fig:adsnoint}(a). Adsorption and desorption proceed at the same rate, see Fig.~\ref{fig:adsnoint}(b). The average number of particles in each cavity at different times are plotted in Figs.~\ref{fig:adsnoint}(c) and \ref{fig:adsnoint}(d), for respectively adsorption ($\nav^{\mathrm{ads}}(x,t)$) and desorption ($\nav^{\mathrm{des}}(x,t)$). In Fig.~\ref{fig:adsnoint}$(f)$ we plot $\nav^{\mathrm{ads}}(x,t) + \nav^{\mathrm{des}}(x,t) - 13$. Since adsorption and desorption proceed at the same rate this quantity is zero for all $x$ and $t$. If adsorption proceeds faster than desorption it is positive, and \textit{vice versa}.

In a continuous system with constant transport diffusion $D_t$, the concentration dependence $c(x,t) = \nav (x,t)$ during desorption can be found by solving the diffusion equation:
\begin{equation}
\frac{\partial c(x,t)}{\partial t} = D_t \frac{\partial^2 c(x,t)}{\partial x^2},
\end{equation}
with boundary conditions 
\begin{align}
c(0,t) &= c(L,t) = 0, \quad \forall t, \\
c(x,0) &= \nav_{\mathrm{start}} , \quad  0 < x < L.
\end{align}
The solution is \cite{crank1979}:
\begin{equation}\label{eq:soldesdtc}
c^{\mathrm{des}}(x,t) = \frac{4 \nav_{\mathrm{start}}}{\pi} \sum_{n = 0}^{\infty} \frac{1}{2 n +1} \sin \left( \frac{\left(2n+1 \right) \pi x}{L} \right) e^{-D_t \left[ (2n+1) \pi  / L  \right]^2 t}.
\end{equation}
The concentration dependence during adsorption is simply $c^{\mathrm{ads}}(x,t) = \nav_{\mathrm{start}} - c^{\mathrm{des}}(x,t)$. It was checked that for $L \uparrow \infty$ our simulations converge to this solution.

\subsection{Repulsive particles}

\begin{figure}
\centering
\resizebox{1.0\columnwidth}{!}{%
\includegraphics{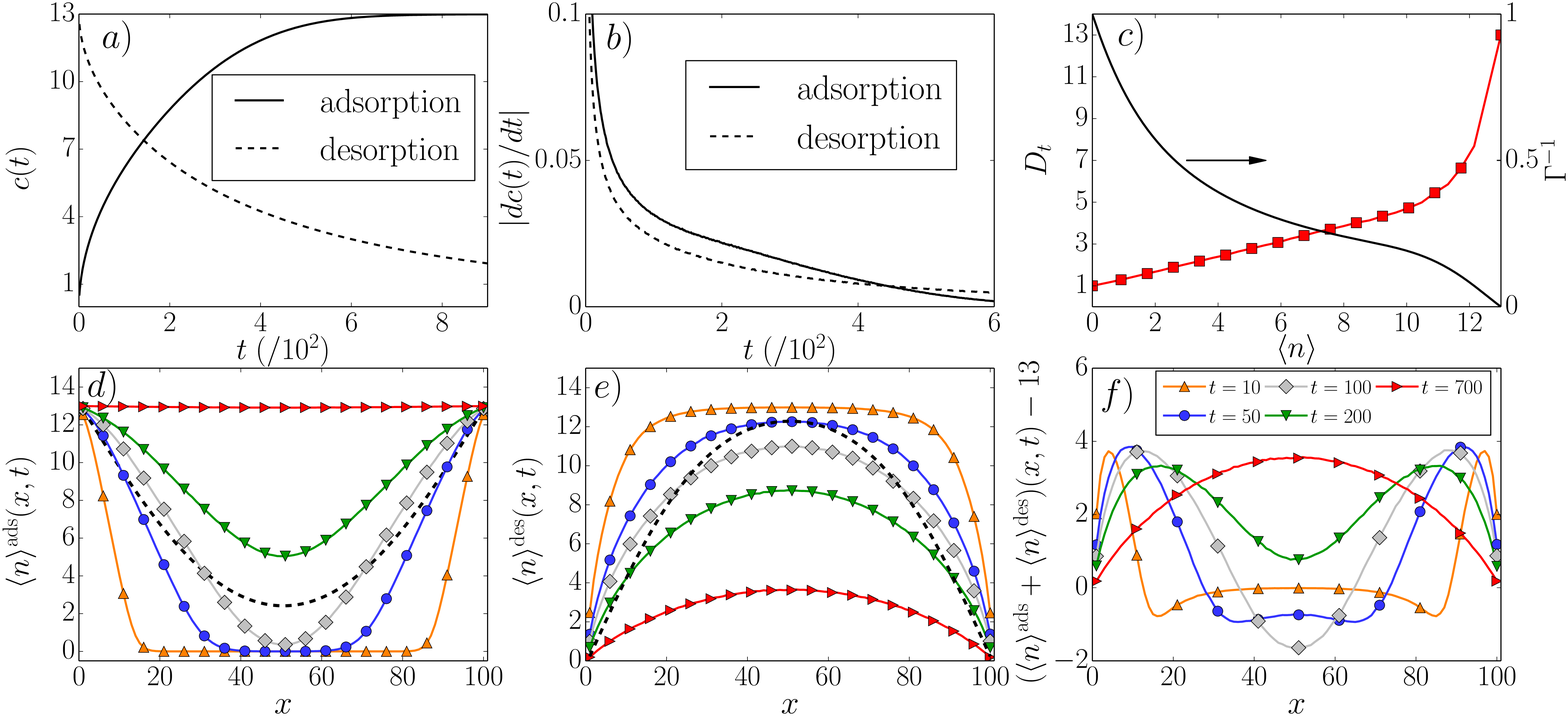} }
\caption{Adsorption/desorption between $\nav = 13$ and $\nav = 0$, for $f(n) = 0.2 n^2$, $\nmax = 13$, and rates Eq.~(\ref{eq:ratesletter}). $a)$ Average concentration $c(t)$. $b)$ Rate of adsorption and desorption. $c)$ (red squares) Transport diffusion from kMC, the line is a guide to the eye. $D_t$ at $\nav = 0$ and $\nav = 13$ was calculated analytically. (black line) $\Gamma^{-1}$ $d), e)$ Average number of particles $\nav(x,t)$ in each cavity at different times $t$, during respectively adsorption $\nav^{\mathrm{ads}}(x,t)$ and desorption $\nav^{\mathrm{des}}(x,t)$. Markers are shown each 5 positions. The lines are a guide to the eye. The black dashed lines represent respectively $\nav^{\mathrm{ads}}(x,t)$ and $\nav^{\mathrm{des}}(x,t)$ for the parameters of Fig.~\ref{fig:adsnoint}, with the same concentration $c(t)$ as for $t=100$ in this figure. $f)$ $\nav^{\mathrm{ads}}(x,t) +\nav^{\mathrm{des}}(x,t) - 13$. Markers are shown each 10 points. The lines are a guide to the eye.}
\label{fig:adsrep}
\end{figure}

\begin{figure}
\centering
\resizebox{1.0\columnwidth}{!}{%
\includegraphics{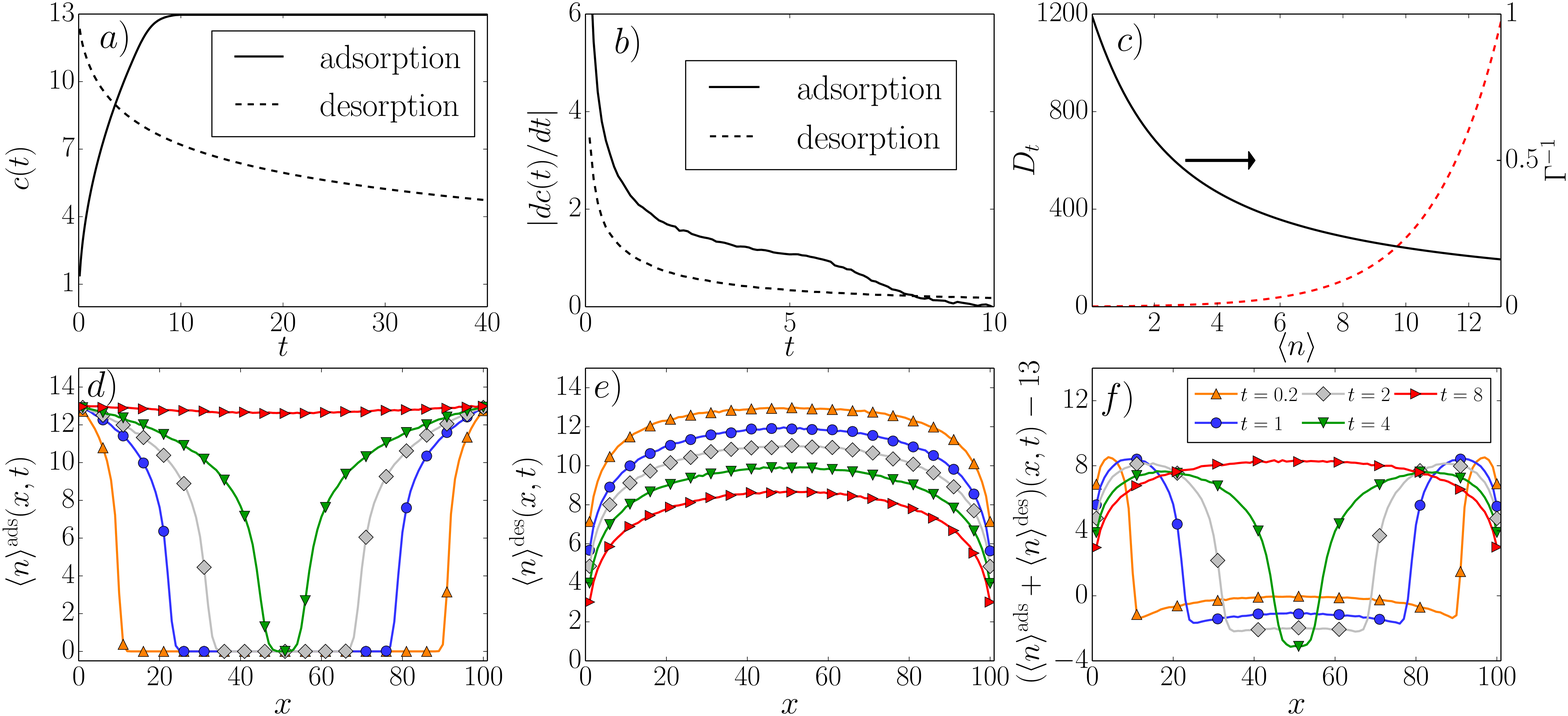} }
\caption{Adsorption/desorption between $\nav = 13$ and $\nav = 0$, for $f(n) = 0.2 n^2$, no $\nmax$, and rates Eq.~(\ref{eq:trzrp}). $a)$ Average concentration $c(t)$. $b)$ Rate of adsorption and desorption. $c)$ (dashed line) Transport diffusion (analytical result). (black line) $\Gamma^{-1}$ $d), e)$ Average number of particles $\nav(x,t)$ in each cavity at different times $t$, during respectively adsorption $\nav^{\mathrm{ads}}(x,t)$ and desorption $\nav^{\mathrm{des}}(x,t)$. Markers are shown each 5 positions. The lines are a guide to the eye. $f)$ $\nav^{\mathrm{ads}}(x,t) +\nav^{\mathrm{des}}(x,t) - 13$. Markers are shown each 10 points. The lines are a guide to the eye.}
\label{fig:adsrepzrp}
\end{figure}

For interacting particles the transport diffusion is concentration dependent and, as a result, adsorption and desorption proceed at different rates \cite{BOOKruthven,CESGarg,SPT2003}. For a continuous system the concentration dependence can be found by solving the diffusion equation with a concentration-dependent transport diffusion:
\begin{equation}\label{eq:desorpgeneral}
\frac{\partial c(x,t)}{\partial t} = \frac{\partial}{\partial x} \left( D_t(c(x,t)) \frac{\partial c(x,t)}{\partial x} \right),
\end{equation}
with the correct boundary conditions for adsorption/desorption. If the transport diffusion can be calculated analytically, as in Section \ref{sec:3}, this equation can be solved numerically. If not, it is necessary to first perform kMC simulations to measure the transport diffusion at different concentrations \cite{PRLbecker}. This result can then be interpolated to obtain $D_t(c)$, which can be used to numerically solve Eq.~(\ref{eq:desorpgeneral}). Such a procedure is however time consuming, and one has to be careful with the numerical accuracy of the obtained result. Instead, we simulate directly the adsorption and desorption behavior using kMC. 

Consider the parameters $f(n) = 0.2 n^2$, $\nmax = 13$, and rates Eq.~(\ref{eq:ratesletter}), with the adsorption and desorption proceeding between $\nav = 0$ and $\nav = 13$, cf.~Fig.~\ref{fig:adsrep}. The reservoirs are put at $\nav = 13$ by taking the chemical potential $\mu \rightarrow \infty$ ($p^{\mathrm{eq}}_n(\infty) = \delta_{n \nmax}$). The system system starts at $\nav = 13$ by taking $n_i = 13$ for $1 < i < L$.
The particles are repulsive for this interaction \cite{PRLbecker,Arxivbecker}. The transport diffusion therefore increases with concentration, cf.~Fig.~\ref{fig:adsrep}$(c)$. The rate of adsorption is higher than the rate of desorption, as can be seen from Fig.~\ref{fig:adsrep}$(b)$. If the system is almost completely filled in the adsorption process the desorption starts proceeding faster. 
$\nav^{\mathrm{ads}}(x,t)$ and $\nav^{\mathrm{des}}(x,t)$ at different times are plotted in respectively Figs.~\ref{fig:adsrep}$(d)$ and \ref{fig:adsrep}$(e)$. In Fig.~\ref{fig:adsrep}$(f)$ we plot $\nav^{\mathrm{ads}}(x,t) + \nav^{\mathrm{des}}(x,t) - 13$. Since the transport diffusion grows with increasing concentration, particles diffuse faster from high to low concentration if the particle concentration is high. During adsorption the reservoirs provide a steady input of particles, which creates a front of high concentration that moves into the system. During desorption the region of high concentration gradually disappears. Adsorption therefore proceeds at a higher rate. The front of high concentration moving into the system during adsorption (Fig.~\ref{fig:adsrep}$(d)$) results in two inward moving peaks in Fig.~\ref{fig:adsrep}$(f)$. For small times, the desorption in the middle of the system is faster than the adsorption. The effect is however smaller, and disappears when the middle of the system decreases in concentration. Consequently, particles diffuse at the same rate. By comparing Figs.~\ref{fig:adsnoint} and \ref{fig:adsrep} one sees that adsorption and desorption proceed faster compared to the non-interacting case, as can be expected. The black dashed lines in Figs.~\ref{fig:adsrep}$(d)$ and $(e)$ represent respectively $\nav^{\mathrm{ads}}(x,t)$ and $\nav^{\mathrm{des}}(x,t)$ for the parameters of Fig.~\ref{fig:adsnoint}, with the same average concentration $c(t)$ as for $t=100$ in Fig.~\ref{fig:adsrep}. During the adsorption of non-interacting particles, a particle diffuses as fast near the boundaries as in the middle. For repulsive particles the diffusion is higher at the boundaries than in the middle of the system. This makes the concentration of repulsive particles higher near the boundaries and lower in the middle, for the same average concentration. During desorption the repulsive particles diffuse faster in the middle of the system, also leading to a concentration that is higher near the boundaries and lower in the middle.

We now consider a ZRP with $f(n) = 0.2 n^2$, with the adsorption and desorption proceeding between $\nav = 0$ and $\nav = 13$, cf.~Fig.~\ref{fig:adsrepzrp}. Since the difference in transport diffusion between $\nav = 0$ and $\nav = 13$ is much higher than in the previous case, the difference between adsorption and desorption is more pronounced. The qualitative behavior stays the same.

\subsection{Attractive particles}

\begin{figure}
\centering
\resizebox{1.0\columnwidth}{!}{%
\includegraphics{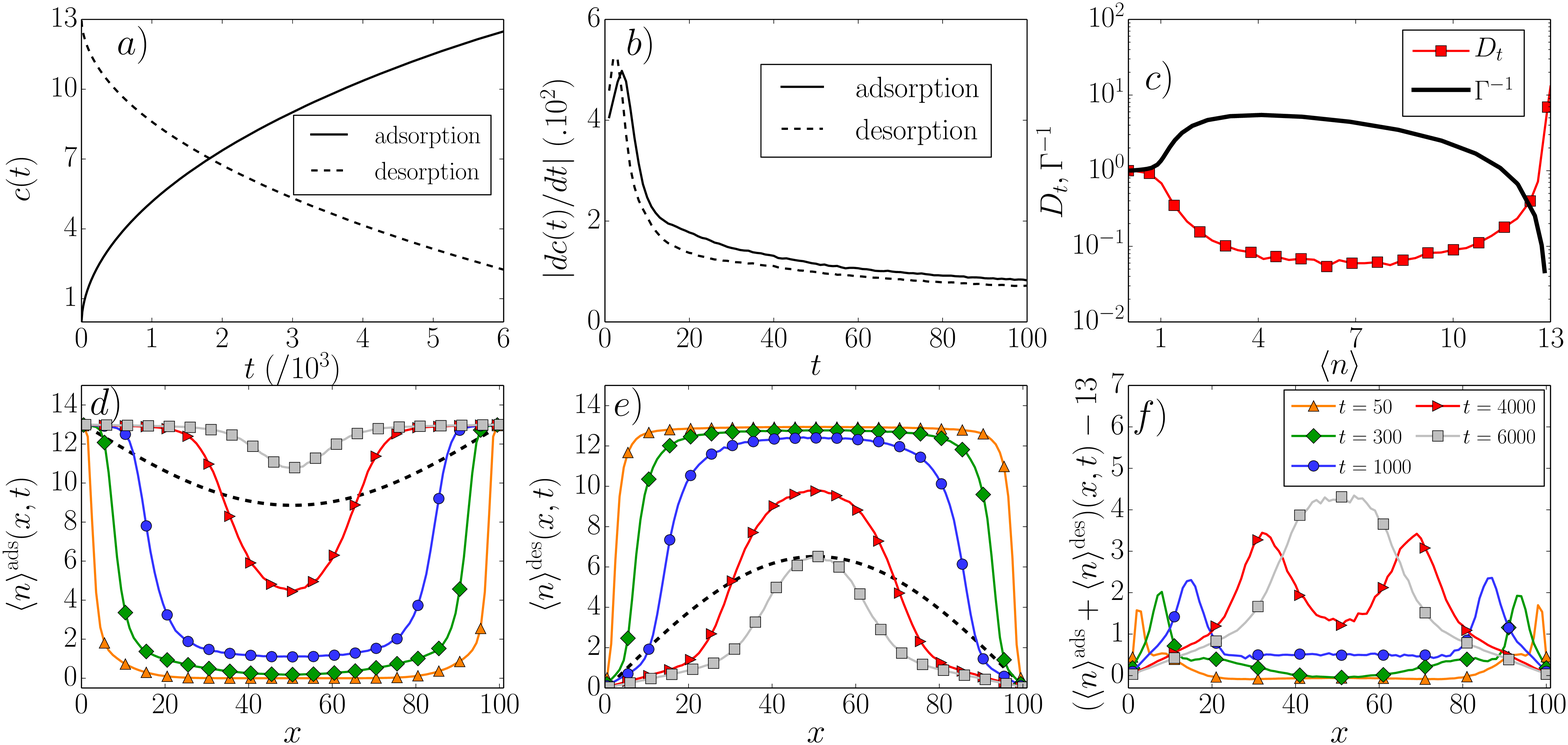} }
\caption{Adsorption/desorption between $\nav = 13$ and $\nav = 0$, for $f(n) =  0.000642 n^2 - 0.0083 n^3$, $\nmax = 13$, and rates Eq.~(\ref{eq:ratesletter}). $a)$ Average concentration $c(t)$. $b)$ Rate of adsorption and desorption. $c)$ (red squares) Transport diffusion from kMC, the line is a guide to the eye. $D_t$ at $\nav = 0$ and $\nav = 13$ was calculated analytically. (black line) $\Gamma^{-1}$ $d), e)$ Average number of particles $\nav(x,t)$ in each cavity at different times $t$, during respectively adsorption $\nav^{\mathrm{ads}}(x,t)$ and desorption $\nav^{\mathrm{des}}(x,t)$. Markers are shown each 5 positions. The lines are a guide to the eye. The black dashed lines represent respectively $\nav^{\mathrm{ads}}(x,t)$ and $\nav^{\mathrm{des}}(x,t)$ for the parameters of Fig.~\ref{fig:adsnoint}, with the same concentration $c(t)$ as for $t=4000$ in this figure. $f)$ $\nav^{\mathrm{ads}}(x,t) +\nav^{\mathrm{des}}(x,t) - 13$. Markers are shown each 10 points. The lines are a guide to the eye.}
\label{fig:adsatt1}
\end{figure}

\begin{figure}
\centering
\resizebox{1.0\columnwidth}{!}{%
\includegraphics{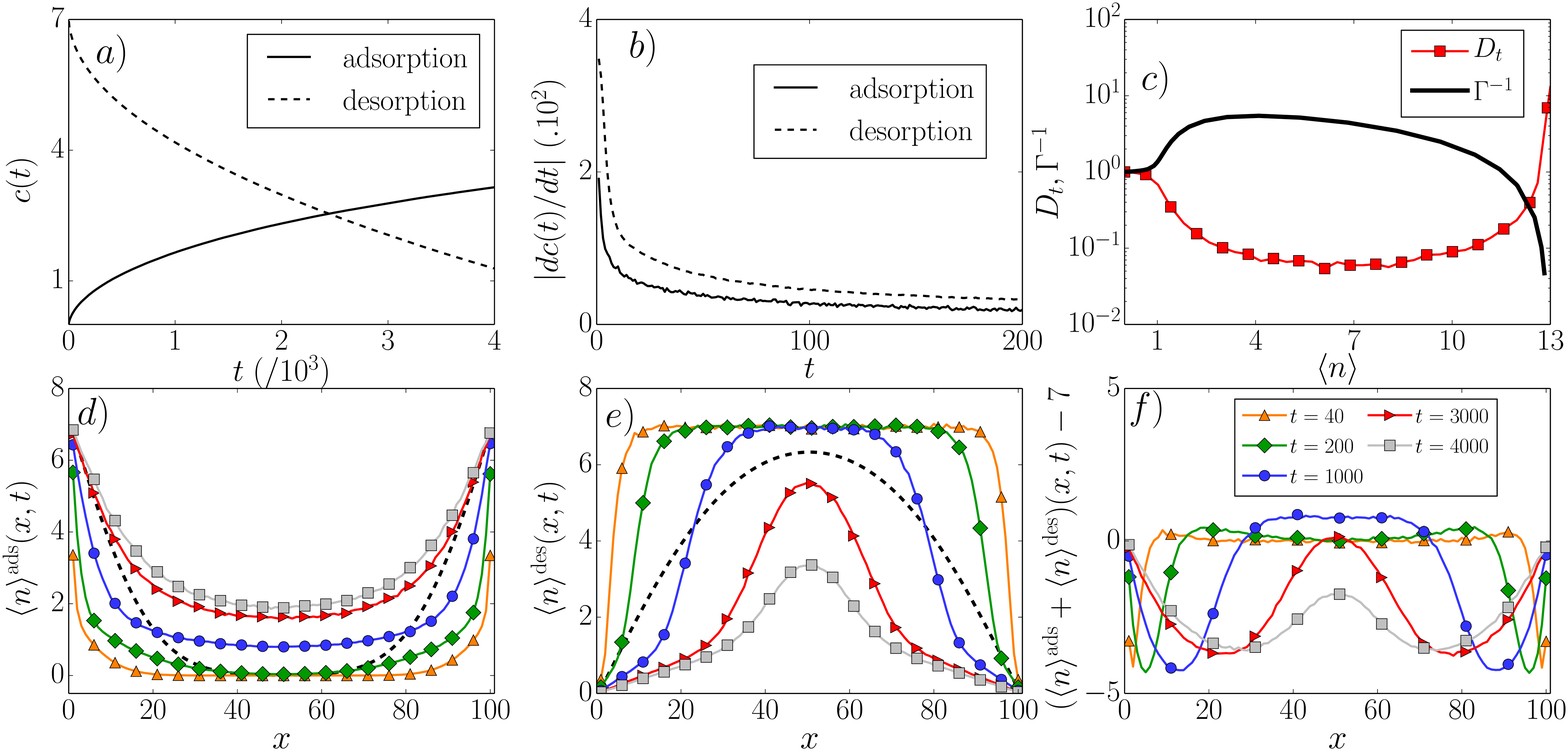} }
\caption{Adsorption/desorption between $\nav = 7$ and $\nav = 0$, for $f(n) =  0.000642 n^2 - 0.0083 n^3$, $\nmax = 13$, and rates Eq.~(\ref{eq:ratesletter}). $a)$ Average concentration $c(t)$. $b)$ Rate of adsorption and desorption. $c)$ (red squares) Transport diffusion from kMC, the line is a guide to the eye. $D_t$ at $\nav = 0$ and $\nav = 13$ was calculated analytically. (black line) $\Gamma^{-1}$ $d), e)$ Average number of particles $\nav(x,t)$ in each cavity at different times $t$, during respectively adsorption $\nav^{\mathrm{ads}}(x,t)$ and desorption $\nav^{\mathrm{des}}(x,t)$. Markers are shown each 5 positions. The lines are a guide to the eye. The black dashed lines represent respectively $\nav^{\mathrm{ads}}(x,t)$ and $\nav^{\mathrm{des}}(x,t)$ for non-interacting particles between $\nav = 7$ and $\nav = 0$, with the same concentration $c(t)$ as for $t=1000$ in this figure. $f)$ $\nav^{\mathrm{ads}}(x,t) +\nav^{\mathrm{des}}(x,t) - 7$. Markers are shown each 10 points. The lines are a guide to the eye.}
\label{fig:adsatt2}
\end{figure}

We now study attractive particles. Consider the parameters $f(n) = 0.000642 n^2 - 0.0083 n^3$, $\nmax = 13$, and rates Eq.~(\ref{eq:ratesletter}). This interaction is qualitatively similar to $f(n) = - 0.2 n^2$, but is more interesting because it provides a good description of methanol diffusion in ZIF-8 \cite{PRLbecker}. The transport diffusion has a minimum for low and medium concentrations and a maximum near $\nav = 13$, cf.~\ref{fig:adsatt1}$(c)$. 
The adsorption and desorption kinetics between $\nav = 13$ and $\nav = 0$ are shown in Fig.~\ref{fig:adsatt1}. Even though the transport diffusion shows a strong minimum for low and medium concentrations, adsorption still proceeds faster than desorption. This is because of the maximum in the transport diffusion around $\nav = 13$, resulting in the same qualitative behavior as in Figs.~\ref{fig:adsrep} and \ref{fig:adsrepzrp}. Adsorption and desorption proceed slower compared to Figs.~\ref{fig:adsnoint} and \ref{fig:adsrep}, because of the minimum in the transport diffusion. The black dashed lines in Figs.~\ref{fig:adsatt1}$(d)$ and $(e)$ represent respectively $\nav^{\mathrm{ads}}(x,t)$ and $\nav^{\mathrm{des}}(x,t)$ for the parameters of Fig.~\ref{fig:adsnoint}, with the same concentration $c(t)$ as for $t=4000$ in Fig.~\ref{fig:adsatt1}. The difference for adsorption is qualitatively the same as in Fig.~\ref{fig:adsrep}$(d)$, although it is more pronounced due to the large difference in the transport diffusion between low and high concentration. For small times the difference in desorption is qualitatively the same as in Fig.~\ref{fig:adsrep}$(e)$. For longer times this behavior is reversed compared to Fig.~\ref{fig:adsrep}$(e)$. Once the middle of the system is no longer at $\nav = 13$, the diffusion near the boundaries (where $\nav \approx 0$) is faster than in the middle of the system.

The adsorption and desorption kinetics between $\nav = 7$ and $\nav = 0$ for the same parameters are shown in Fig.~\ref{fig:adsatt2}. 
In contrast to the previous cases, the transport diffusion at the starting concentration of the adsorption ($\nav = 7$) is smaller than at $\nav = 0$. The steady flow of particles from the reservoirs now slows down the adsorption compared to the desorption. 
The behavior in Fig.~\ref{fig:adsatt2}$(f)$ is the reverse of the previous cases. The black dashed lines in Figs.~\ref{fig:adsatt2}$(d)$ and $(e)$ represent respectively $\nav^{\mathrm{ads}}(x,t)$ and $\nav^{\mathrm{des}}(x,t)$ for non-interacting particles between $\nav = 7$ and $\nav = 0$, with the same concentration $c(t)$ as for $t=1000$ in Fig.~\ref{fig:adsatt2}. The difference in adsorption is reversed compared to Fig.~\ref{fig:adsrep}$(d)$: the concentration at the boundaries is here lower than for the non-interacting case. The difference in desorption is the same as for Fig.~\ref{fig:adsatt1}$(e)$ at long times.

For attractive particles with rates Eq.~(\ref{eq:trzrp}) and $\nmax = \infty$ there is particle condensation, as mentioned in Section \ref{sec:ZRP}. We therefore don't study this situation.

\section{Conclusions}\label{sec:conclude}

To conclude, we have analyzed a one-dimensional lattice model that describes diffusion in confined geometries. The transport and self-diffusion can be calculated analytically for length 1, which represents the uncorrelated solution for any length. For certain parameter values the model reduces to a zero-range process, which can be solved analytically for all lengths. The solution is however always uncorrelated. Systems of length 2 with $\nmax =2$ are solved analytically for all interactions, and include correlation effects. In this case correlations always lower the self-diffusion. The transport diffusion is sometimes enhanced by correlations. 
We studied the adsorption and desorption kinetics for different interactions. In the adsorption process the system is initialized at concentration $c_{\mathrm{low}}$. At time $t=0$ the system is connected to particle reservoirs at higher concentration $c_{\mathrm{high}}$, after which equilibration to $c_{\mathrm{high}}$ occurs. The desorption process proceeds reversely: the system is initialized at $c_{\mathrm{high}}$, and the reservoirs are fixed at concentration $c_{\mathrm{low}}$. Both adsorption and desorption processes are strongly influenced by the concentration-dependent transport diffusion. For repulsive particles the transport diffusion is a monotonic increasing function of concentration. In this case adsorption is always faster than desorption. For attractive particles the transport diffusion is nonmonotonic. Around $\langle n \rangle \approx 0$ it decreases for increasing concentration, has a minimum at an intermediate concentration and increases up to its maximal value at $\langle n \rangle = \nmax$. In this situation both adsorption or desorption can proceed faster than the other, depending on the choice of $c_{\mathrm{high}}$ and $c_{\mathrm{low}}$.

\begin{acknowledgement}
This work was supported by the Flemish Science Foundation (FWO-Vlaanderen). The computational resources and services used in this work were provided by the VSC (Flemish Supercomputer Center), funded by the Hercules Foundation and the Flemish Government -- department EWI.
\end{acknowledgement}

%\bibliography{cite_EPJST_Becker}
%\bibliographystyle{epj.bst}

\end{document}